\documentclass[]{aa} 
\usepackage{graphics,txfonts,longtable}  
\usepackage{rotating}

\newcommand{\ms}{m\,s$^{-1}$}
\newcommand{\kms}{km\,s$^{-1}$}
\newcommand{\bz}{$\langle B_{\rm z} \rangle$}
\newcommand{\bs}{$\langle B \rangle$}

\newcommand{\vald}{{VALD}}
\newcommand{\dream}{{DREAM}}

\newcommand{\firps}[1]{\resizebox{\hsize}{!}{\rotatebox{-90}{\includegraphics{#1}}}}
\newcommand{\fifps}[2]{\centering\resizebox{#1}{!}{\includegraphics{#2}}}
\newcommand{\firrps}[2]{\resizebox{#1}{!}{\rotatebox{-90}{\includegraphics{#2}}}}
\def\i{\,{\sc i}} \def\ii{\,{\sc ii}} \def\iii{\,{\sc iii}}
\def\hd{HD\,24712}
\def\mo{{\sc MOST}}

\begin{document}

\title{Pulsation in the atmosphere of the roAp star \hd%
\thanks{Based on observations collected at the Canada-France-Hawaii Telescope (CFHT), at the Nordic Optical 
Telescope (NOT), at the European Southern Observatory, Paranal, Chile, (DDT-274.D-5011), at the Telescopio 
Nazionale Galileo (TNG), and from \mo, a Canadian Space Agency mission operated jointly by Dynacon, Inc., 
the University of Toronto Institute of Aerospace Studies, and the University of British Columbia, 
with assistance from the University of Vienna.}}
\subtitle{I. Spectroscopic observations and radial velocity measurements}

\author{T. Ryabchikova\inst{1,2},
M. Sachkov\inst{2}, W.W. Weiss\inst{1}, T. Kallinger\inst{1}, O. Kochukhov\inst{3},  S. Bagnulo\inst{4}, I. 
Ilyin\inst{5},  J.D. Landstreet\inst{6}, F. Leone\inst{7}, G. Lo\,Curto\inst{3}, T. L\"uftinger\inst{1}, 
D. Lyashko\inst{8},\and  A. Magazzu\inst{9}
        }

\titlerunning{Pulsation in the atmosphere of the roAp star \hd}
\authorrunning{T. Ryabchikova et al.}
\offprints{~\\ T. Ryabchikova at \email{ryabchik@inasan.ru}}

\institute{Institut f\"ur Astronomie, Universit\"at Wien, T\"urkenschanzstrasse 17, 1180 Wien, Austria
      \and Institute of Astronomy, Russian Academy of Sciences, Pyatnitskaya 48, 119017 Moscow, Russia
      \and Department of Astronomy and Space Physics, Uppsala University Box 515, SE-751 20 Uppsala, Sweden
      \and European Southern Observatory, Casilla 19001, Santiago 19, Chile
      \and Astrophysikalisches Institut Potsdam, An der Sternwarte 16, D-14482 Potsdam, Germany
      \and Department of Physics and Astronomy, University of Western Ontario, London, Ontario N6A 3K7, Canada
      \and INAF - Osservatorio Astrofisico di Catania, Via S. Sofia 78, 95123 Catania, Italy
      \and Tavrian National University, Simferopol, Ukraine
      \and INAF - Telescopio Nazionale Galileo, PO Box 565, 38700 Santa Cruz de La Palma, Spain
          }

\date{Received / Accepted }

\abstract {} {We have investigated the structure of the pulsating 
atmosphere of one of the best studied rapidly oscillating Ap stars, \hd.} 
{For this purpose we analyzed spectra collected during 2001\,--\,2004. An extensive data set was obtained in 
2004 simultaneously with the photometry of the Canadian MOST mini-satellite. 
This allows us to connect directly atmospheric dynamics observed as radial velocity variations 
with light variations seen in photometry. } 
{We directly derived for the first time and for different chemical elements, respectively ions, phase shifts 
between photometric and radial velocity pulsation maxima 
indicating, as we suggest, different line formation depths in the atmosphere. 
This allowed us to estimate for the first time the propagation velocity of a pulsation wave 
in the outer stellar atmosphere of a roAp star to be slightly lower than the sound speed.
We confirm large pulsation amplitudes (150\,--\,400\,\ms) for REE lines and the H$\alpha$ core, while 
spectral lines of the other elements (Mg, Si, Ca, and Fe-peak elements) have nearly constant velocities. 
We did not find different pulsation amplitudes and phases for the lines of rare-earth
elements before and after the Balmer jump, which supports the hypothesis of REE concentration in the upper
atmosphere above the hydrogen line-forming layers.
We also discuss radial velocity amplitudes and phases measured for individual spectral lines as tools 
for a 3D tomography of the atmosphere of \hd.} 
{} 

\maketitle

\section{Introduction}
\label{intro} About 10\,\% to 20\,\% of upper main sequence stars are characterized by remarkably rich line 
spectra, often containing numerous unidentified features. Compared to the solar case, overabundances of up to a 
few dex are often inferred for some iron peak and rare earth elements, whereas some other chemical elements are 
found to be underabundant (Ryabchikova~et~al. 2004). Some of these {\it Chemically Peculiar} (CP) stars also 
exhibit organized magnetic fields with a typical strength of a few kG. The specific chemical peculiarities 
observed are believed to result from the influence of the magnetic field on the diffusing ions, possibly in 
combination with the influence of a weak, magnetically directed wind (e.g., Babel 1992).

More than 30 cool CP stars exhibit an additional peculiarity, which is high-overtone, low-degree, non-radial 
$p$-mode pulsation with periods in the range of 6\,--\,21 minutes, with their observed pulsation amplitudes 
modulated according to the visible magnetic field structure. These so-called rapidly oscillating peculiar A to 
F-type (roAp) stars are key objects for asteroseismology, which presently is the most powerful tool for testing 
theories of stellar structure and evolution. Spectroscopic and photometric techniques provide information on 
the boundary zone relevant for any pulsation model, and open access to different modes and hence atmospheric layers. 
An observed phase lag between luminosity and radial velocity variations is an important parameter for a first step 
towards modeling the stellar structure. The dependency of radial velocity amplitudes as a function of optical depths 
lead to a 3D tomography of the stellar atmosphere.

The best studied multi-periodic roAp star presently is \hd, which makes this star a cornerstone for stellar seismology, 
even beyond the class of CP stars. It was discovered to be a pulsator by Kurtz (\cite{kurtz82}) with periods around 
6 minutes, and  Matthews~et~al. (\cite{MWWY88}) found synchronized radial velocity variations. Photometry with the 
Whole Earth Telescope (WET, Kurtz~et~al. \cite{kurtz02}) revealed a `missing' mode, suggesting that $p$-mode pulsation 
are strongly affected by the global stellar magnetic field, an aspect which was investigated in detail by 
Cunha~et~al. (\cite{cunha03}) and by Cunha (\cite{cunha06}).

These characteristics made \hd\ a very strong candidate for contemporaneous spectroscopic observations with large ground 
based equipment suited to obtain high time resolution, high spectral resolution, and high signal-to-noise ratio spectra 
simultaneously with high precision photometric observations with MOST, the Canadian photometric space telescope 
(Walker~et~al. \cite{WMK2003}). The \mo\ instrument is a 15-cm Maksutov type optical telescope feeding twin CCD 
detectors through a broadband filter (350\,--\,700\,nm). The equipment was designed to obtain rapid photometry of 
bright stars for up to 2 months and with a nearly 100\% duty cycle. Despite its low mass of only 54\,kg (and hence little inertia) 
it is able to perform optical photometry of point sources due to a pointing accuracy of better than $\pm 1$\arcsec\,rms  
(Walker~et~al. \cite{WMK2003}).

\begin{table*}
\caption{Journal of time-resolved spectroscopy of \hd. Listed are, among others, the duration of a continuous set of observations (Run, in hours) in a given night and the number of individual spectra taken during such a run, and the typical S/N ratio for the continuum.}
\label{tbl1}
\begin{center}
\begin{tabular}{cccccccccccl}
\noalign{\smallskip}
\hline
\hline
Set& Civil date&Start HJD  &  Spectral range & Run  &No. of & Exposure & Overhead & Typical & Rot. & Instr. \\ 
No. & (UT) &(245 0000+)& (\AA)          & (hr) &spectra& time (s)& time (s) & $S/N$   &phase & \\ 
\hline
\,~1 &2001.10.02 &2185.15654&6106-6189&3.77 &123&60&42&100&0.71 &GECKO\\ 
\,~2 &2001.10.03 &2186.15303&6106-6189&4.60 &162&60&41&100&0.79 &GECKO\\ 
\,~3 &2001.10.04 &2187.15230&6620-6730&4.77 &163&60&42&100&0.87 &GECKO\\ 
\,~4 &2002.09.23 &2541.08000&5822-5887&2.44 &\,~82&60&44&100&0.28 &GECKO\\ 
\,~5 &2002.09.24 &2541.16417&6543-6658&1.43 &\,~49&60&44&100&0.29 &GECKO\\ 
\,~6 &2002.09.24 &2542.02444&5822-5887&2.03 &\,~66&60&44&100&0.36 &GECKO\\ 
\,~7 &2002.09.25 &2542.11958&5284-5344&1.96 &\,~66&60&44&100&0.37 &GECKO\\ 
\,~8 &2002.09.25 &2543.05228&5822-5887&1.89 &\,~65&60&44&100&0.44 &GECKO\\ 
\,~9 &2002.09.26 &2543.14946&6105-6195&1.84 &\,~64&60&44&100&0.45 &GECKO\\ 
10 &2002.09.26 &2544.14997&5822-5887&3.81 &127&60&44&100&0.53 &GECKO\\
11 &2003.11.06 &2949.69312&4540-9952&1.58 &\,~53&50&55&\,~80&0.08 &SOFIN\\ 
12 &2004.11.11 &3320.78693&3850-6730&2.31 &\,~92&60&30&120&0.87 &HARPS\\ 
13 &2004.11.12 &3321.74421&3400-6720&2.09 &\,~92&50&30&300&0.94 &UVES\\  
14 &2004.11.13 &3322.77598&3400-6720&1.73 &\,~73&50&30&300&0.03 &UVES\\  
15 &2004.11.15 &3324.60032&4575-7872&1.07 &\,~35&60&52&120&0.18 &SARG\\  
16 &2004.12.02 &3341.66789&4575-7872&1.13 &\,~33&60&52&120&0.55 &SARG\\  
\hline
\end{tabular}
\end{center}
\end{table*}

\mo\ observed \hd\ continuously from Nov. 5, to Dec. 4, 2004, and a parallel ground based observing campaign was organized 
which yielded the spectroscopic time series listed in the last five lines of Table\,\ref{tbl1}. While the main photometric 
results will be published elsewhere, we focus here on the spectroscopic analysis and are using \mo\ data primarily for a 
direct comparison of the data taken simultaneously in space and from ground.

\section{Observations and spectra reduction}
\label{obser}

The observations of \hd\ were collected during 13 nights: Oct. 2-4, 2001; Sept. 23-26, 2002; 
Nov. 6, 2003; Nov. 11-13, 15, and  Dec. 2, 2004. The journal of observations is given in Table\,\ref{tbl1}, which 
lists set numbers, civil dates, heliocentric Julian dates of the start of the observing sequence, spectral range, 
run duration in hours, and the number 
of spectra that were obtained in each night. The chosen exposure times are a compromise between the requirement to 
integrate the spectrum only over a small fraction of the pulsation period, and the need to have a reasonable signal-to-noise 
($S/N$) ratio for each spectrum.  The seventh column gives the rotation phases for the mean time of each data set according 
to the ephemeris given by Ryabchikova~et~al. (\cite{Period}):

\hspace{3mm}HJD(\bz$_{max})=2453235.18(40)+12.45877(16)$ d.

\noindent
Heliocentric Julian dates are given for the centre of exposures. 
The heliocentric corrections were applied to spectroscopic observations and to \mo\ photometry in 
the same way.

\subsection{Time-series of single-order spectra}

\hspace{2mm}$\bullet$~~{\em GECKO}: 
The observations of 2001 and 2002 were obtained with the single-order $f/4$ GECKO coud\'e spectrograph and the EEV1-CCD 
at the 3.6-m Canada-France-Hawaii telescope. The spectra have a resolving power of about 115\,000, determined from the 
widths of a number of ThAr comparison lines. The exposure time was 60\,s, dead time was 44\,s, and the achieved $S/N$ 
in the continuum was about 100. These observations covered 7 spectral regions centered approximately at 4860, 5300, 5855, 
6160, 6600, 6675, and 7780\,\AA, containing the most interesting spectral lines of singly and doubly-ionized rare earth 
elements (REE), H$\alpha$, H$\beta$, O\i, Fe\i, Ca\i, and Ba\ii.

The spectra were reduced using standard IRAF tasks. Each stellar, flat and calibration frame had a mean bias subtracted 
and was then cleaned of cosmic ray hits and collapsed to one dimension. The spectra were divided by a mean flat-field, 
extracted in the same way, and the continuum was fitted with a three-segment cubic spline, using the same rejection parameters 
for all spectra so that the continuum fit is as uniform as possible. 

The wavelength scale was established using about 40 lines of a ThAr emission lamp, resulting in an rms scatter of about 
$3\times 10^{-4}$\,\AA. The wavelength scale was linearly interpolated between ThAr lamp spectra taken before and after 
the stellar series, but the spectra were not sampled to a linear wavelength spacing. 

\subsection{Time-series of \'echelle spectra }

\hspace{2mm}$\bullet$~~{\em SOFIN}:
The observations from Nov. 6, 2003, were carried out with the SOFIN high resolution \'echelle spectrograph at the 2.56\,m 
Nordic Optical Telescope (NOT), La Palma, Spain. Each spectrum had an integration time of 50\,s with a readout time 
of 55\,s, giving a time-resolution of 105\,s. The typical $S/N$ ratio is about 80 and the resolving power 
$\approx$\,80\,000. These \'echelle spectra cover the region from 5000 to 6800\,\AA. The \'echelle images 
were reduced with the Advanced Acquisition, Archiving, and Analysis (4A) package written in C (Ilyin \cite{II00}).

The standard reduction sequence includes bias subtraction from the CCD overscan, photon noise estimation for the pixel 
variances, correction for the CCD fixed pattern noise using a master flat field (a sum of 100 exposures), subtraction 
of the scattered light determined from a 2D spline fit to the inter-order gaps. The spectral order position is found 
from the flat field image and subsequently adjusted for each \'echelle science frame. This step is followed by a 
weighted extraction of spectral orders with elimination of cosmic spikes based on a linear regression. The shape 
of the spectra and fringes in the red part of the CCD are corrected with a flat field spectrum smoothed with a spline fit. 
The wavelength calibration is based on about 1\,300 ThAr spectral lines collected from two successive images, using a 
2D fit to them, taking also the time of exposures for calibration and science frames into account. A zero point 
correction had to be applied which resulted in a final RV error of about 25\,\ms\ at the image center.

\hspace{2mm}$\bullet$~~{\em HARPS\,~{\rm and}\,~UVES}:
The 2004 spectroscopic observation were carried out with HARPS (High Accuracy Radial velocity Planet Searcher) spectrometer 
at the 3.6-m telescope at ESO, La Silla. 92 spectra with 60\,s exposure time, S/N\,=\,120, and 120\,000 resolving power 
were taken during November 10/11, 2004, simultaneously with MOST. Because of the unique coincidence with the space 
photometry, Director's Discretionary Time (274.D-5011) was granted for November 11/12 and 12/13 with the UVES 
spectrograph at the 8.2-m telescope, UT2 (Kueyen), of the VLT on Paranal, Chile, (92 \& 73 spectra, 50\,s exposure time, 
S/N\,=\,300, with a resolving power of about 80\,000).

All spectra were reduced and normalized to the continuum level with a routine specially developed by one of us (DL) 
for a fast reduction of time-series observations. It is a component of the spectral reduction package STAR\,XP, a 
special modification of the Vienna automatic pipeline for \'echelle spectra processing (Tsymbal~et~al. \cite{tsymbal}). 
All bias and flat field images were median averaged before calibration. The scattered light was subtracted by using a 
2D background approximation. For cleaning of cosmic rays we used a new algorithm which compares the direct and reversed 
spectral profiles. To determine the spectrum order boundaries, the code uses a special template for each order position 
in each row across the dispersion axis. The shift of the row spectra relative to the template is derived by a 
cross-correlation technique. Wavelength calibration was done by the usual 2D fit. The accuracy of this procedure is 
$\approx$20\,\ms. The final step of continuum normalization was done by transforming of the flat field blaze function 
to the response function in each order.

\hspace{2mm}$\bullet$~~{\em SARG}: 
During \mo\ observations of \hd, additional spectra were obtained in Nov. 14/15, 2004, (35 spectra), 
and on Dec. 01/02, 2004, (33 spectra), with the high resolution spectrograph (SARG) at the 3.55-m 
{\it Telescopio Nazionale Galileo} (TNG) at the Observatorio del Roque de los Muchachos (La Palma, Spain). 
The spectra were reduced using standard ESO-MIDAS software with the same main steps as described above. 
The spectra cover the range of 4570\,--\,7900\,\AA, have a resolving power of about 57\,000 and a S/N 
ratio of approximately 120. The time resolution was 129\,s (60\,s for exposure and 69\,s for read out).

\subsection{Polarimetry with SOFIN}
\label{mag-obs}

The spectropolarimetric observations of \hd\ were carried out between Oct. 29, and Nov. 18, 2003, with the high resolution 
\'echelle spectrograph, SOFIN, attached to the Cassegrain focus of NOT. The spectrograph is equipped with three different 
cameras offering three different resolving powers. To obtain observations in the polarimetric mode, the second camera with 
a resolving power of $\approx$\,80\,000 was used. Between 4000\,--\,7000\,\AA\ seven spectral orders, each covering about 
40 to 50\,\AA\ were used for the magnetic field analysis. 

\begin{table}
\caption{Journal of spectropolarimetric observations of \hd. The longitudinal field \bz\ was estimated using ten Nd {\sc ii} and Nd {\sc iii} lines (3rd column), and seven Cr {\sc i}, Cr {\sc ii}, and Fe {\sc i} lines (4th column).}
\label{tbl2}
\begin{center}
\begin{tabular}{c|c|rr}
\hline
\hline
   HJD     & Rotation    & \multicolumn{2}{c}{\bz\ (G)} \\ 
(245 0000+)&   phase     &Nd {\sc ii}, Nd {\sc iii}&Cr {\sc i}, Cr {\sc ii}, Fe {\sc i}  \\
\hline
2941.6516 & 0.44  & $ 598\pm 165$&$  87\pm ~~68$\\
2943.6341 & 0.60  & $ 909\pm 154$&$ 240\pm 132$\\ 
2945.5758 & 0.76  & $1109\pm 105$&$ 535\pm 145$\\ 
2946.5973 & 0.84  & $1090\pm 144$&$ 729\pm 180$\\ 
2947.6067 & 0.92  & $1182\pm 100$&$1098\pm ~~98$\\
2948.6514 & 0.00  & $1132\pm ~~98$&$1064\pm 130$\\
2952.6267 & 0.32  & $ 835\pm 174$&$ 258\pm ~~79$\\
2953.6572 & 0.40  & $ 629\pm 129$&$ 155\pm ~~86$\\
2954.6316 & 0.48  & $ 720\pm 192$&$ 165\pm 155$\\ 
2955.6350 & 0.56  & $ 830\pm 187$&$ 137\pm 107$\\ 
2956.5954 & 0.64  & $1062\pm 178$&$ 260\pm 174$\\ 
2957.5958 & 0.72  & $1159\pm 164$&$ 483\pm 158$\\ 
2961.6160 & 0.04  & $1168\pm 100$&$1048\pm 152$\\ 
\hline
\end{tabular}
\end{center}
\end{table}

The circularly polarized spectra were obtained with a Stokesmeter, consisting of a fixed achromatic quarter-wave plate, 
a beam splitter made of a calcite plate, and an achromatic rotating quarter-wave plate, whose position is controlled by 
a stepping motor. To obtain accurate circular polarization measurements, usually a sequence of four exposures is obtained. 
Each of the beams is exposed twice, with the quarter-wave plate rotated by $90\degr$ after the first and before the last 
exposure. Such a sequence reduces instrumental effects to a minimum, because in the images taken with the quarter-wave 
plate rotated by $90\degr$, instrumental signatures change sign and cancel when averaging the two exposures. 

Data reduction was performed with the aforementioned 4A software package including all standard procedures, such as bias 
subtraction, flat field correction, subtraction of the scattered light, weighted extraction of the orders, and bad 
pixel (cosmic ray) corrections. ThAr exposures obtained before and after each observing night were used to perform 
wavelength calibration and to test for possible spurious instrumental polarization, caused e.g. by bending of the 
spectrograph which is directly mounted on the telescope, different positions of the star on the slit, or temporal 
variations of the seeing. S/N ratios for the observed spectra are typically 200\,--\,300. Rotation phases of 
\hd\ (see Table~\ref{tbl2}) were calculated according to the ephemeris and rotation period derived by 
Ryabchikova~et~al. (\cite{Period}).

\section{Radial velocities and magnetic field strength}
\label{rvmag}

For radial velocity (RV) measurements we carefully chose unblended or minimally blended lines in the
3300\,--\,6800\,\AA\ spectral region. Between 3900 and 4400\,\AA\ the cores of the strong (resonance) lines of Ca, Fe
and Sr were measured. Our choice was based on synthetic spectrum calculations over the whole spectral region of
3300\,--\,6800\,\AA, made with the spectral synthesis code {\sc synth3} written by Kochukhov, and using the atmospheric
parameters and abundances from Ryabchikova~et~al. (\cite{RL97}). Atomic parameters of spectral lines for the  synthesis
were extracted from the Vienna Atomic Line Database, \vald\ (Kupka~et~al. \cite{VALD2}),  and from the Database for Rare
Earths at Mons University, \dream\ (Bi\'emont~et~al. \cite{dream99}), which is also accessible via the \vald\ extraction
procedures. For the Nd\iii\ identification additional atomic data from Crosswhite (\cite{Crswt}), Aldenius
(\cite{Ald01}) and Ryabchikova~et~al. (\cite{Nd3}) were used.

The radial velocities were measured with a center-of-gravity technique and attention was payed to the stability of 
the spectrographs. HARPS time-resolved spectra provide stable results with a mean rms dispersion of 20\,\ms\ per 
individual non-pulsating line, while a quasi-linear, long-term drift was found in both nights of the observations 
with UVES. These drifts were approximated with a smooth spline function based on the average measurements of a few 
unblended non-pulsating Fe\i\ lines. This quasi-linear drift was then subtracted from the RV measurements of all 
other spectral lines. It should be emphasized that the instrumental variation of the spectrograph's zero point 
occurs on a much longer time scale than the stellar {\it p-}mode variability, and therefore does not affect the 
pulsation analysis presented here.

Radial velocities of more than 500 unblended spectral lines were measured in the spectral region from 3900 to 
6800\,\AA, and additional 80 lines were measured in the region blueward the Balmer jump (BJ). A complete list of 
measured lines (but not all individual measurements) together with their identification is given in Table\,\ref{puls} (Online material). 
The purpose of this table is to provide line identifications and to indicate pulsating and non-pulsating lines.

\begin{figure}[!t]
\firps{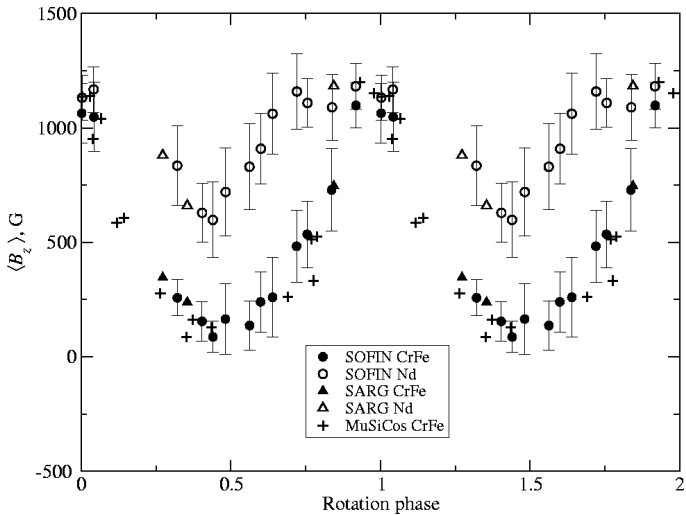}
\caption{Longitudinal magnetic field variations in \hd. Open (Nd lines) and filled (Cr-Fe lines) circles are the  observations 
presented in this
paper, open and filled triangles are magnetic measurements from Leone \& Catanzaro (\cite{LC04}), and crosses represent
data taken from Ryabchikova~et~al. (\cite{Period}).}
\label{mf}
\end{figure}

The longitudinal magnetic field was measured as the first moment of the observed Stokes V parameter for a set of chosen
spectral lines. We have performed separate measurements of the iron-peak elements (Cr and Fe) and of the REEs (Nd and
Tb). The results and error estimates are given in Table\,\ref{tbl2} and are illustrated in Fig.\,\ref{mf}. For a
comparison, \bz\ data from Ryabchikova~et~al. (\cite{Period} -- MuSiCoS) and from Leone \& Catanzaro (\cite{LC04} -- SARG) 
are also included in this figure.

We found that some spectral lines with large Land\'e factors are partially split in non-polarized spectra. In particular
one line, Cr\i\,5247.56\,\AA, is a pure triplet with $g_{\rm eff}$\,=\,2.51, and another line, Fe\ii\,6432.48\,\AA, 
is a pseudo-doublet with $g_{\rm eff}$\,=\,1.82. Using these lines we estimated the magnetic modulus \bs\, at phases 
0.867, 0.944 (close to the magnetic maximum) and at phase 0.42 (near the magnetic minimum). For the latter, a UVES spectrum 
of \hd\ was extracted from the ESO archive. \bs\ estimates were made by fitting  calculated synthetic line profiles 
to the observed spectra. Magnetic synthetic calculations were carried out with \mbox{SYNTHMAG} 
(Kochukhov \cite{K06}), which represents an improved version of the program described by Piskunov (\cite{P99}). 
\bs\ varies between 3100\,--\,3300\,G at \bz$_{max}$ and  2500\,G at \bz$_{min}$ according to our estimates.

\section{Frequency analysis}
\label{freq}

\begin{figure}[!t]
\centering \fifps{8.5cm}{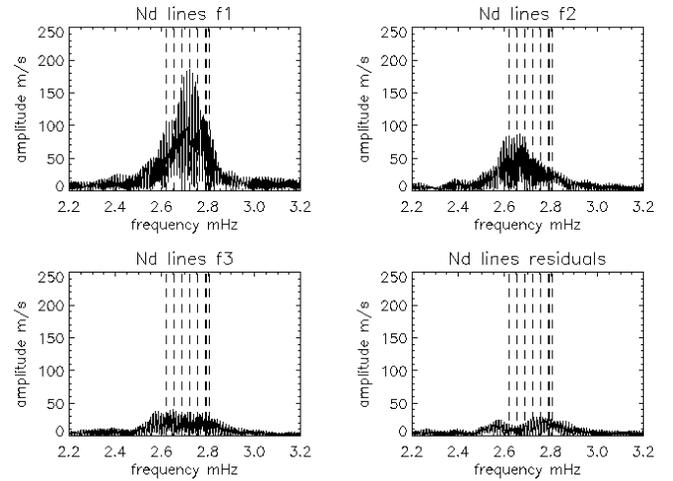}
\caption{Amplitude spectra of the Nd\iii\ spectral lines of the 2004 observations. The top left panel shows the DFT of the 
original RV data; the top right panel represents the DFT after prewhitening with the highest amplitude frequency; the other panels show the next prewhitening steps. Dashed lines indicate photometric frequencies  according to  Kurtz at al. (\cite{kurtz05}). } 

\label{dft}
\end{figure}

Although RV variations in the REE lines due to pulsation are very distinctive and the relative accuracy of individual 
spectroscopic data is higher than for photometry, it proved to be difficult to study in detail the frequencies of multiperiodic 
roAp stars by spectroscopy, because a large telescope is needed during an extended period of time. For a reliable frequency 
analysis it is necessary to observe continuously during weeks with a minimum of gaps. This is possible either with dedicated satellites, 
as is MOST for photometry, or with multisite ground based campaigns, such as WET (Kurtz~et~al. \cite{kurtz05}). Although our spectroscopic 
monitoring does not allow for a detailed frequency analysis, we performed nevertheless such an analysis to confirm the consistency of 
the main frequencies in the spectroscopy obtained simultaneously with the MOST photometry.

\begin{figure*}
\fifps{15.4cm}{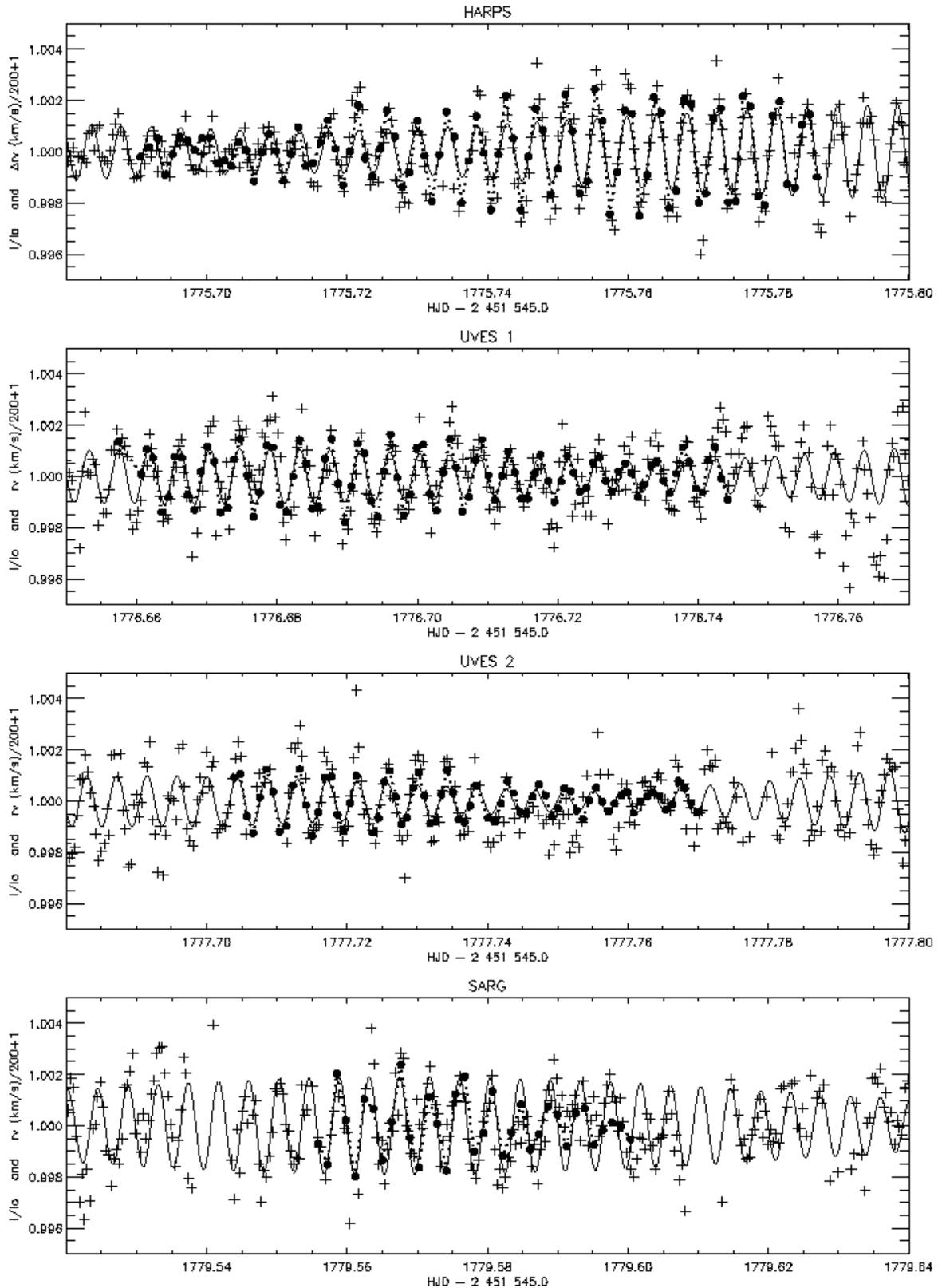}
\caption{Normalized RV variations for Pr\iii\ (filled circles) are compared with a synthetic light curve (black solid line,   
computed with the three largest amplitude frequencies derived from the entire \mo\ observing run) and with simultaneous \mo\ photometric data (crosses). 
The dotted line connects the Pr\iii\,5299\,\AA\ spectral line RV values (dots) and it follows well the solid line, based on \mo\
photometry.}
\label{shift}
\end{figure*}

\begin{figure*}[t]
\centering \fifps{11cm}{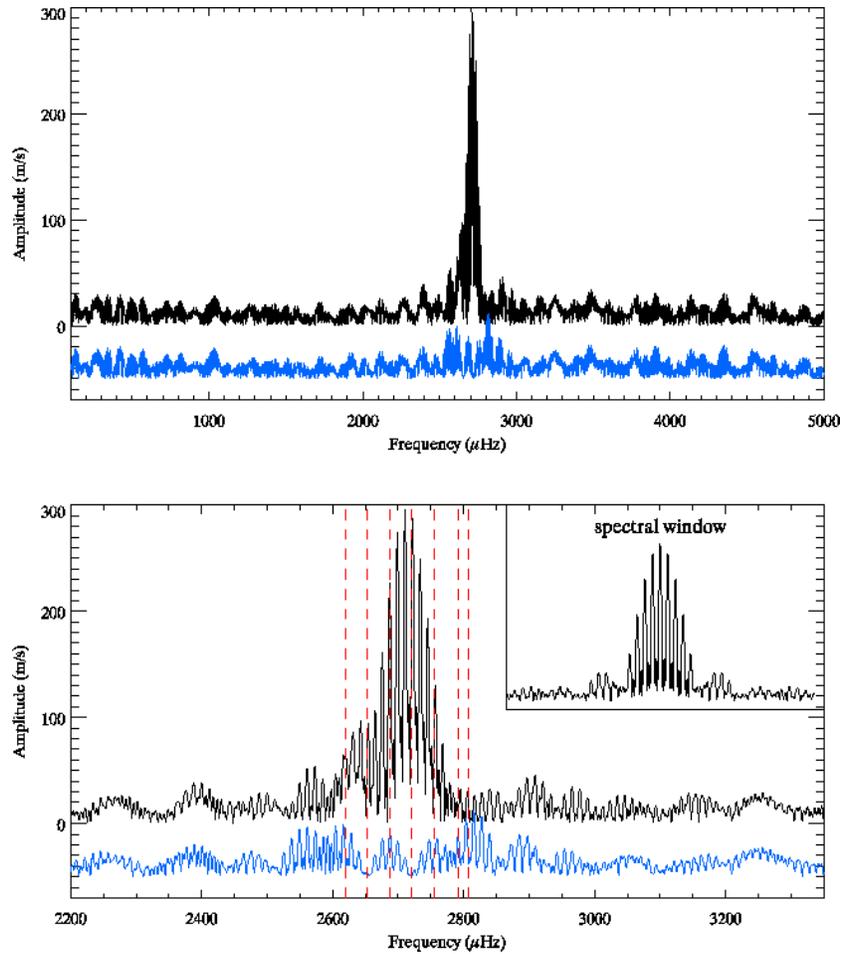}
\caption{Amplitude spectra of Nd\iii\ spectral lines observed in 2001. 
In each panel the upper curve shows the original amplitude spectrum, whereas the lower curve
(shifted downwards for better visibility)
represents the Fourier transform of the RV-values after prewhitening with the two main RV frequencies of 
2720.96 and 2652.96 $\mu$Hz, which account for most of the RV power in the data set. 
The lower panel represents an enlarged view of the upper one. The spectral window of the data set is inserted.
The vertical dashed lines indicate photometric frequencies according to Kurtz at al. (\cite{kurtz05}).} 
\label{dft_2001}
\end{figure*}

Our Fourier analysis of the RV data of selected lines -- Pr\iii\ 5284, 5300\,\AA, Nd\iii\ 5203,
5294, 5845, 5851, 5987, 6145\,\AA\ and Tb\iii\,5505\,\AA\ --  was based on a discrete Fourier
transform (DFT) and stepwise prewhitening with a sine fit to the highest amplitudes 
(see Fig.\,\ref{dft}). In this analysis we used all spectra obtained simultaneously with \mo\ during four nights
around the magnetic maximum (sets 12 to 15). The duty cycle for this combined 4-night data set is 
poor (about 8\%), but knowing from \mo\ photometry which alias to avoid, we found the 3 highest
amplitude periods in our spectroscopy to be 6.125\,min, 6.282\,min, and 6.202\,min.   These values
agree well with contemporaneous MOST photometry, as is illustrated in Figure\,\ref{shift} for one of
the pulsating  lines (Pr\iii\,5300\,\AA). These frequencies correspond to $\nu_{4}$, $\nu_{2}$, and
$\nu_{3}$ in Kurtz at al. (\cite{kurtz05}). The frequency analysis performed for the 2001
Nd\iii\ data gives two main frequencies $\nu_{4}$ and $\nu_{2}$ (identification according to Kurtz et al., op.cit.) with similar amplitudes as in
2004 (see Fig.~\ref{dft_2001} in Online Material). This figure illustrates (as does Fig.\,\ref{dft}) the possibility to identify pulsation frequencies, amplitudes and phases even in short spectroscopic runs distributed over several nights -- what results in a very poor duty cycle -- provided one can avoiding aliases thanks to \mo\ and WET photometry.

The short observing run in 2003 does not allow us to
resolve frequencies. Only one frequency close to $\nu_{3}$ was derived.

Different authors prefer to characterize periodic signal either with periods or with frequencies. For convenience of the reader  
we mention here the conversion: period in minutes transforms to a frequency in mHz via $\nu$(mHz)\,=\,16.6667/P(min).

\section{Phase relations between photometry and spectroscopy}
\label{phase}

Our time-resolved observations in 2004 were carried out simultaneously with the Canadian micro--satellite MOST which monitored 
\hd\ from 2004 November 6, to December 5. This provides us with the opportunity to derive directly the phase lag between 
photometry and spectroscopy. 

This was attempted already earlier by Matthews~et~al. (\cite{MWWY88}) despite a rather poor S/N ratio ($\approx$\,20) of their 
individual spectra. But the large number of spectra ($\approx$\,600) allowed the authors to derive an average RV curve and to find a 
coincidence of RV maxima with $\delta$B minima, which corresponds to a phase lag of about 0.5, where we define a phase lag
as a phase difference between 
the maxima of both, photometric and RV variations. They found the photometric maximum to occur typically $after$ the RV maximum, which we
confirm with our data.

The high S/N and spectral resolution of the present observations resulted in more precise values for phase lags, 
and in particular it allowed us to determine this quantity for individual spectral lines. In order to minimize the influence of 
the higher point-to-point scatter of the photometric data with respect to the spectroscopic observations, we computed 
an artificial time-series data set based on the 3 dominant frequencies and their amplitudes and phases 
($\nu_2$, $\nu_3$ and $\nu_4$) which were derived from the complete set of \mo\ observations. 
Next, this artificial time-series 
was cross-correlated with the RV observations of the individual spectral lines. The time interval for the cross correlation 
ranges from plus to minus 6.125 minutes, the latter corresponds to the period with the largest amplitude ($\nu_4$). 
The time step was 1 second. The best correlation gives the time lag between photometry and spectroscopy expressed in seconds. 

The time lags obtained with artificial time-series data generated with the 3 dominant frequencies differ only by
$\sim$2\,sec from those obtained with a full set of frequencies which is about an order of magnitude less than the
accuracy of the used correlation technique. The remaining frequencies in the full \mo\ frequency solution with very low
amplitudes obviously do not affect the time lag determination.

In Fig.\,\ref{shift} we illustrate the excellent agreement between the RV variations of the Pr\iii\ spectral line and 
the photometric observations, shifted by about -197\,sec, which corresponds to a phase lag of -0.54, using the main 
photometric pulsation frequency. Both observations are normalized and scaled for better visibility.

The result of the cross-correlation procedure is given in Table\,\ref{most}. The brightness maximum occurs for all lines after 
the RV maximum, but the phase lag itself depends strongly on the individual line. It is largest for the H$\beta$ line, which 
has a minimal RV amplitude of 91\,\ms, and gradually decreases for lines showing higher amplitudes. It will be shown in a 
following paper that this gradual change in phase lag is probably connected with line formation in the atmosphere.

\begin{table}
\caption{Phase lag in seconds between luminosity and RV variations for different chemical species. The fourth column
gives the same phase lags based on a pulsation period of 6.125\,min.}
\label{most}
\begin{center}
\begin{tabular}{lccc}
\hline
\hline
Line    &  $\lambda$ (\AA) & \multicolumn{2}{c}{ Phase lag   } \\
        &                   &in seconds      & in periods \\
\hline   
H$\beta$ & 4861 &  -356$\pm$25 & -0.97$\pm$0.08 \\  
Eu\ii\   & 6645 &  -356$\pm$22 & -0.97$\pm$0.06 \\  
Nd\ii\   & 6650 &  -313$\pm$22 & -0.85$\pm$0.06 \\  
H$\alpha$& 6563 &  -307$\pm$22 & -0.84$\pm$0.06 \\  
Nd\iii\  & 5286 &  -301$\pm$21 & -0.82$\pm$0.06 \\  
Nd\ii\   & 5255 &  -297$\pm$21 & -0.81$\pm$0.06 \\  
Nd\iii\  & 6690 &  -294$\pm$22 & -0.80$\pm$0.06 \\  
Nd\iii\  & 5851 &  -283$\pm$22 & -0.77$\pm$0.06 \\  
Dy\iii\  & 5730 &  -278$\pm$22 & -0.76$\pm$0.06 \\  
Nd\iii\  & 5845 &  -270$\pm$21 & -0.73$\pm$0.06 \\  
Nd\iii\  & 5203 &  -255$\pm$21 & -0.69$\pm$0.06 \\  
Nd\iii\  & 5294 &  -247$\pm$22 & -0.67$\pm$0.06 \\  
Pr\iii\  & 5300 &  -197$\pm$21 & -0.54$\pm$0.06 \\  
Tb\iii\  & 5505 &  -104$\pm$22 & -0.28$\pm$0.06 \\  
\hline				     
\end{tabular}			     
\end{center}
\end{table}

\section{Radial velocity variations of individual elements}

For all measured lines we did a period search with the periodogram method. This analysis allowed us to crudely
estimate the probability that a given period is true (Horne \& Baliunas \cite{horne}), They developed an algorithm which applies to fully resolved frequency spectra. 
The results -- RV amplitudes with 
the error, period and error, and probability of the period -- are given in Table\,\ref{puls} (Online Material). HJD\,=\,2453320.0 
was chosen as a reference time for our pulsation analysis. 
It is remarkable that the weighted mean from all periods with a probability higher than 0.99 determined from individual lines 
yields exactly the value of the most prominent photometric period observed in 2000\,--\,2004, which is P\,=\,6.125\,min.

The large number of measurable lines of different chemical species allows for a detailed analysis of pulsation waves in the 
atmosphere of \hd, but which will be presented in a follow up paper. Here we discuss briefly pulsation properties derived 
for different chemical elements.

\begin{figure*}[]
\centering
\firrps{82mm}{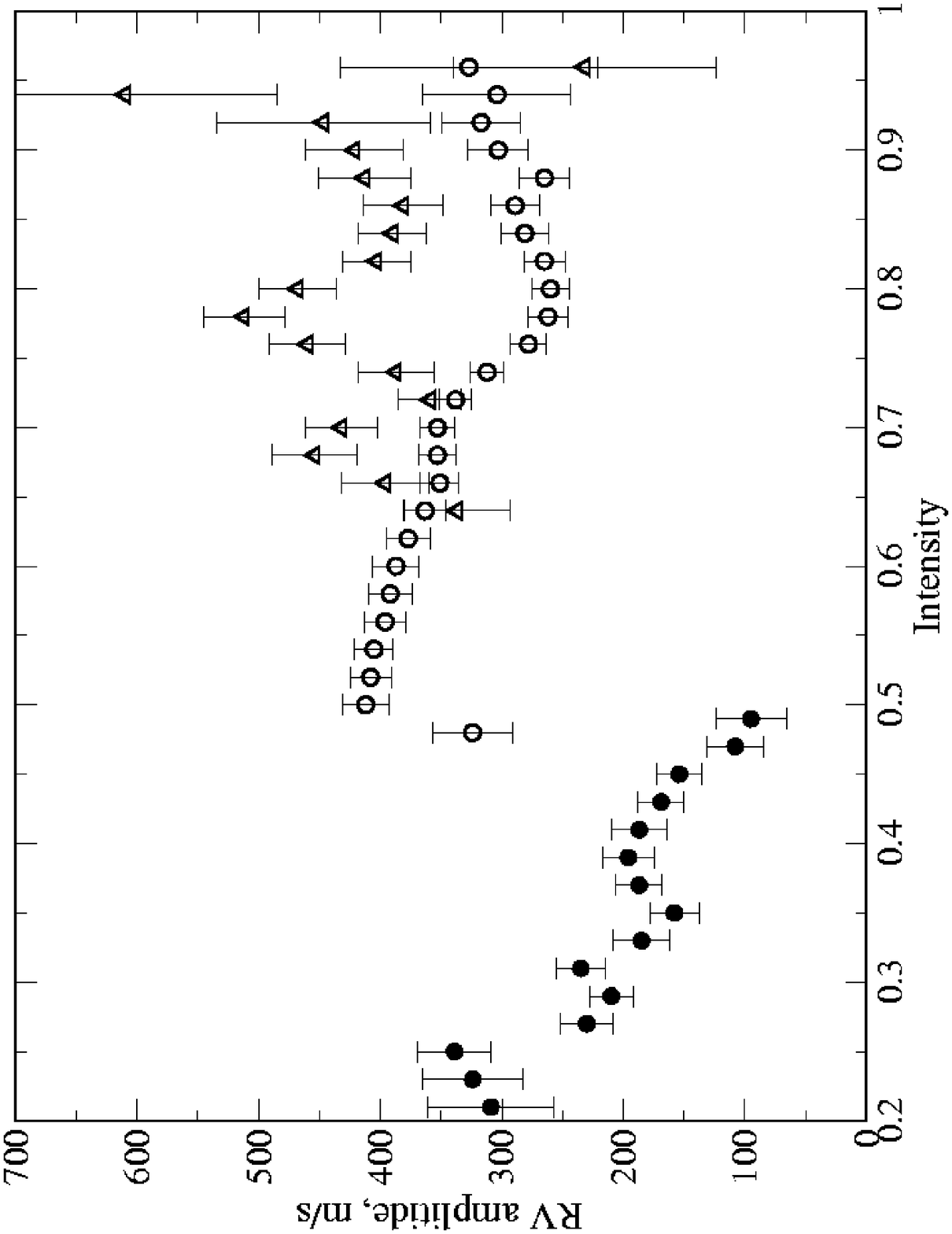}\hspace{0.5cm}\firrps{82mm}{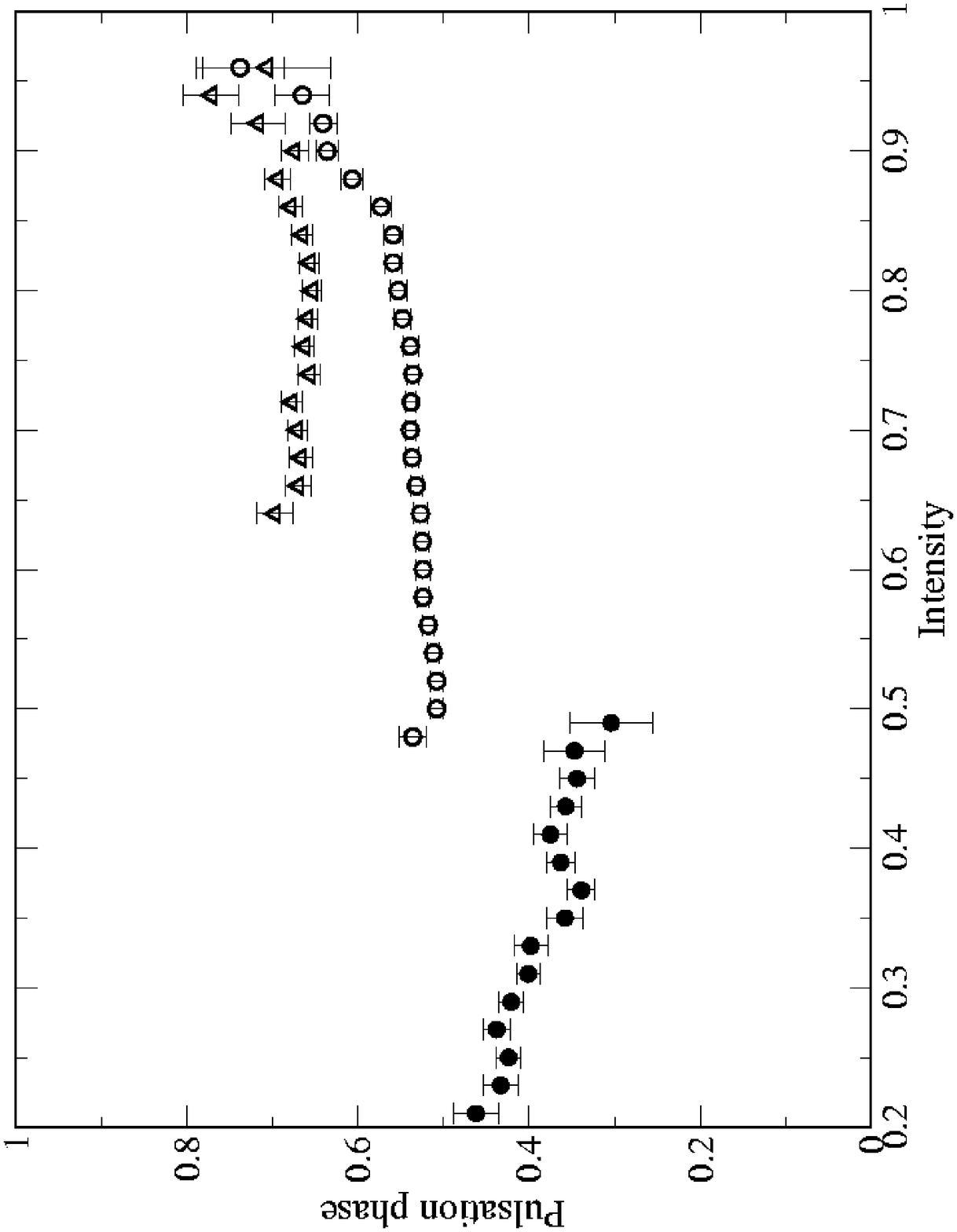}
\caption{Bisector measurements of the H$\alpha$ line (filled circles), lines of Nd\iii\,5294\,\AA\ (open circles),
 Pr\iii\,5300\,\AA\ (open triangles). 
 (crosses). The RV amplitudes are shown in the left panel and pulsation phases (based on P\,=\,6.125\,min) in the right panel.}
\label{bisector1}
\end{figure*}

\hspace{0mm}$\bullet$~~{\em Hydrogen:}
H$\gamma$, H$\beta$ and H$\alpha$ cores indicate pulsation with the amplitude increasing from H$\gamma$ to H$\alpha$.
Our measurements support the results obtained by Balona \& Laney (\cite{alcir}) for $\alpha$\,Cir and by Balona (\cite{hr3831}) for
HR\,3831, and all together they provide a direct evidence for the growth of pulsation amplitudes towards the upper atmospheric layers.
Bisector measurements of the H$\alpha$ line are shown in
Fig.\,\ref{bisector1} (RV amplitudes -- right panel, phases -- left panel). Both, amplitudes and phases increase with
line depth, i.e. towards the upper atmospheric layers as is observed also in other roAp stars (for example
$\alpha$\,Cir, Baldry~et~al. \cite{acir}; $\gamma$\,Equ, Sachkov~et~al. \cite{sach04a}; or HD\,99563, Elkin et al. \cite{EKM05}).

\begin{figure}[]
\centering
\firrps{82mm}{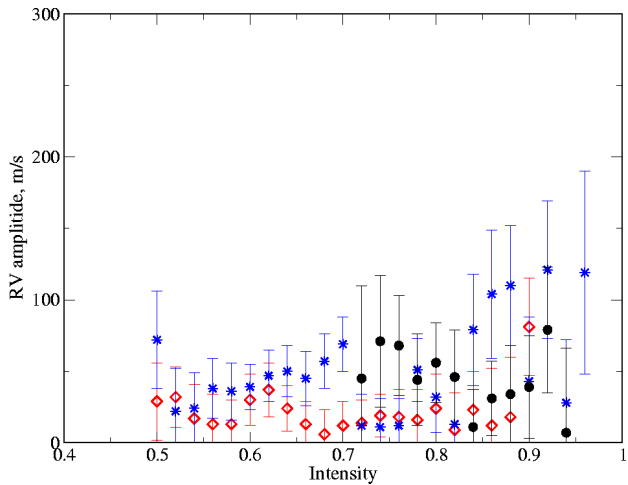}
\caption{Bisector RV amplitudes (based on P\,=\,6.125\,min) of Fe\ii\,5169\,\AA\ (open diamonds), Mg\i\,5172\,\AA\ (asterisks), and Ca\i\,6122\,\AA\ 
 (filled circles).}
\label{bisector2}
\end{figure}

\hspace{0mm}$\bullet$~~{\em Na, Mg, S and Si:} Lines of these elements do not reveal any pulsation. The upper limit for 
the RV amplitudes ranges from 10 to 30\,\ms, depending on the line strength and hence accuracy of measurements. Bisector 
measurements across the Mg\i\,5172\,\AA\ line (Fig.\,\ref{bisector2}) also rejects variability with an amplitude above 40\,\ms. 
With the exception of the bisector close to the continuum (r\,=\,0.96) we never recover the true pulsation period in a 
periodogram. The pulsation signal at r\,=\,0.96 is explained by a La\ii\ blend in the red wing of the  Mg\i\ line. 
A significant pulsation amplitude 
was also measured in the 4696.20\,\AA\ line, which coincides with a S\i\ feature, but which may be attributed to an unclassified 
Nd\iii\ line at $\lambda$\,4696.205\,\AA. The latter information is from the unpublished lists of Crosswhite (\cite{Crswt}) 
which were the main source for official NIST data on Nd\iii\ energy levels (Martin~et~al. \cite{NBS-REE78}).

\hspace{0mm}$\bullet$~~{\em Fe-peak elements:} About half of the measured Ca lines show oscillations compatible with the 
photometrically observed pulsation periods. In four cases we found a signal with 96\% significance, among which is the core 
of the resonance Ca\ii\,3933\,\AA\ line, obviously formed high in the atmosphere. Bisector measurements across the 
Ca\i\ 6122.22\,\AA\ line are shown in Fig.\,\ref{bisector2}. Although this line indicates weak variation with a period 
of 6.125\,min (and a probability of 0.82\%), the bisector variations do not differ from those of
constant Mg\i\ and Fe\ii\ lines. A small unknown blend of a REE would be enough to produce a spurious, very low 
amplitude variation in even a strong, but non-pulsating line (see below).

Three out of 5 Sc\ii\ lines show pulsation, but they all are blended with REE lines. Similarly, nine out of 21 measurable Ti 
lines show small amplitude variations with the known pulsation period. The Ti\ii\,4501.26\,\AA\ line is blended with 
Nd\iii\,4501.23\,\AA\ which contributes to 
about 25\% of the total line intensity, thus resulting in the pulsation signal with the typical phase of Nd lines.

\begin{figure}[]
\centering
\firps{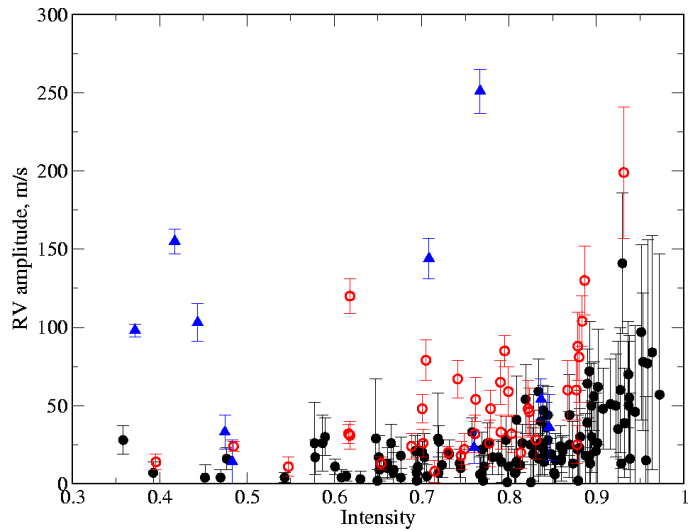}
\caption{Pulsation amplitude versus central residual intensity for lines of Fe-peak elements.
Filled/open circles indicate lines with no pulsation/known pulsation period, respectively. Fe-peak 
lines known to be blended with the REEs are marked with triangles.}
\label{Fe-peak}
\end{figure}

Cr, Mn, Fe, and Co lines do not pulsate. Only 25\% of the whole set of measured lines have  pulsation periods typical for \hd, 
and part of these lines are blended with lines of REEs. Figure\,\ref{Fe-peak} displays RV amplitudes derived from sine-fits 
to lines of Fe-peak elements with a pulsation period of 6.125\,min, and as a function of central residual intensities. 
We can conclude the absence of pulsation with amplitudes exceeding 15\,\ms\ in the whole atmospheric range where Fe-peak lines 
are formed. An apparent increase of the RV amplitude for weaker lines simply reflects a reduced accuracy of the measurements.

A few lines of the Fe-peak elements have rather large RV amplitudes (triangles in Fig.\,\ref{Fe-peak}) and in most
cases this is a result of blending with a REE. Two spectral lines at 5208\,\AA\ and 5429\,\AA\ are of particular
interest, because they are usually identified as Cr\i\ and Fe\i\ lines, respectively, and their pulsation
characteristics attributed to these elements (e.g., Elkin~et~al. \cite{EKM05}). Actually, these lines are heavily blended with 
Pr\iii\,5208\,\AA\ and Nd\iii\,5429\,\AA, respectively, and consequently show typical REE pulsation phases, but with reduced amplitudes.

\hspace{0mm}$\bullet$~~{\em Sr to Ba:} Spectral lines of Sr, Y, Zr, Rh, Pd, In and Ba were measured in the spectrum of \hd. Five 
Ba\ii\ lines are constant to within 10\,\ms, as are also lines of Rh\i\ and Pd\i. A spectral feature at $\lambda$\,4511.26\,\AA, 
identified as In\i\,4511.31\,\AA, may be blended with an unclassified line of Ho (Crosswhite \cite{Crswt}). Pulsation is seen 
in Sr\ii\ lines and in five out of 11 lines of Y\ii. We carefully checked for blends and can exclude this possibility as explanation 
for a pulsation signal. Pulsation does not appear in weaker lines of Y\ii\ and there is a definite dependence of the RV amplitude 
and phase on line intensity. It seems that Y\ii\ lines originate in the atmosphere where lines with high pulsation amplitudes just 
start to be formed.

\hspace{0mm}$\bullet$~~{\em Rare Earth elements:} We have measured 260 lines of 13 REEs in the first and the second
ionization stage. Almost all of them show pulsation with large amplitudes and different pulsation phases, depending on
the species and line intensity (see also Table\,\ref{most}). Bisector measurements at different (continuum normalized) intensity 
levels of two representative lines,
Nd\iii\,5294\,\AA\ and Pr\iii\,5300\,\AA, are shown in Fig.\,\ref{bisector1}. Although RV variations 
are present, they are not as large as in the roAp star $\gamma$\,Equ (Sachkov~et~al. \cite{sach04a}). No difference 
in the pulsation signature is found for lines of the same element/ion located on both sides of the Balmer jump.
For example, two Nd\iii\ spectral lines, $\lambda$ 3603 and $\lambda$ 6145\,\AA, formed at approximately the same depth in
the stellar atmosphere according to Mashonkina~et~al. (\cite{MRR05}) have RV amplitudes of 
185 and 194\,\ms, and pulsation phases of 0.74 and 0.79, respectively (data set 13, UVES).

\hspace{0mm}$\bullet$~~{\em Thorium:} Two lines identified as Th\iii\ were measured, but no significant pulsation was
detected. Using equivalent width measurements and the model atmosphere from Ryabchikova~et~al. (\cite{RL97}), we
estimate a Th abundance of $\log(Th/N_{\rm tot})$\,=\,$-9.26\pm0.12$. Oscillator strengths for Th\iii\ lines were taken
from Bi\'emont~et~al. (\cite{Th3}). The thorium abundance in \hd\ is comparable to that in HD\,101065 (Cowley~et~al. \cite{HD101065}), 
and the thorium overabundance in the atmosphere is similar to the overabundance  of most REE obtained from the first
ions. Note, that Th abundance in both stars, \hd\ and in HD\,101065 has been derived using partition functions (PF) from
Kurucz' {\sc ATLAS9} code.

\hspace{0mm}$\bullet$~~{\em Unidentified lines:} Along with lines of well established identification we measured all 
unidentified features with equivalent width $\ge$10\,m\AA. The total number of these lines is 115, and about one third of 
them coincide with the position of  Nd\iii\ lines from Crosswhite's unpublished list. Because they are not yet classified, 
we consider them as unidentified lines requiring a proper identification. Most of these potential Nd\iii\ lines have pulsation 
phases in the range of 0.4 to 0.5, corresponding well to the classified Nd\iii\ lines. Only one line at $\lambda$\,4748.17\,\AA\ 
does not show pulsation variations, all other lines reveal pulsation with the typical amplitudes and phases of REEs. We conclude 
that plenty of still unknown REE lines are present in the spectra of roAp stars.

The ability to constrain classification of unidentified lines on the basis of their pulsation 
amplitudes and phases and thus provide useful information for laboratory studies is worth mentioning. 
This is a unique property of roAp stars, and it was already used in the classification
study of the Nd\iii\ lines (Ryabchikova~et~al. \cite{Nd3}).
  
\begin{figure*}[!th]
\fifps{8.5cm}{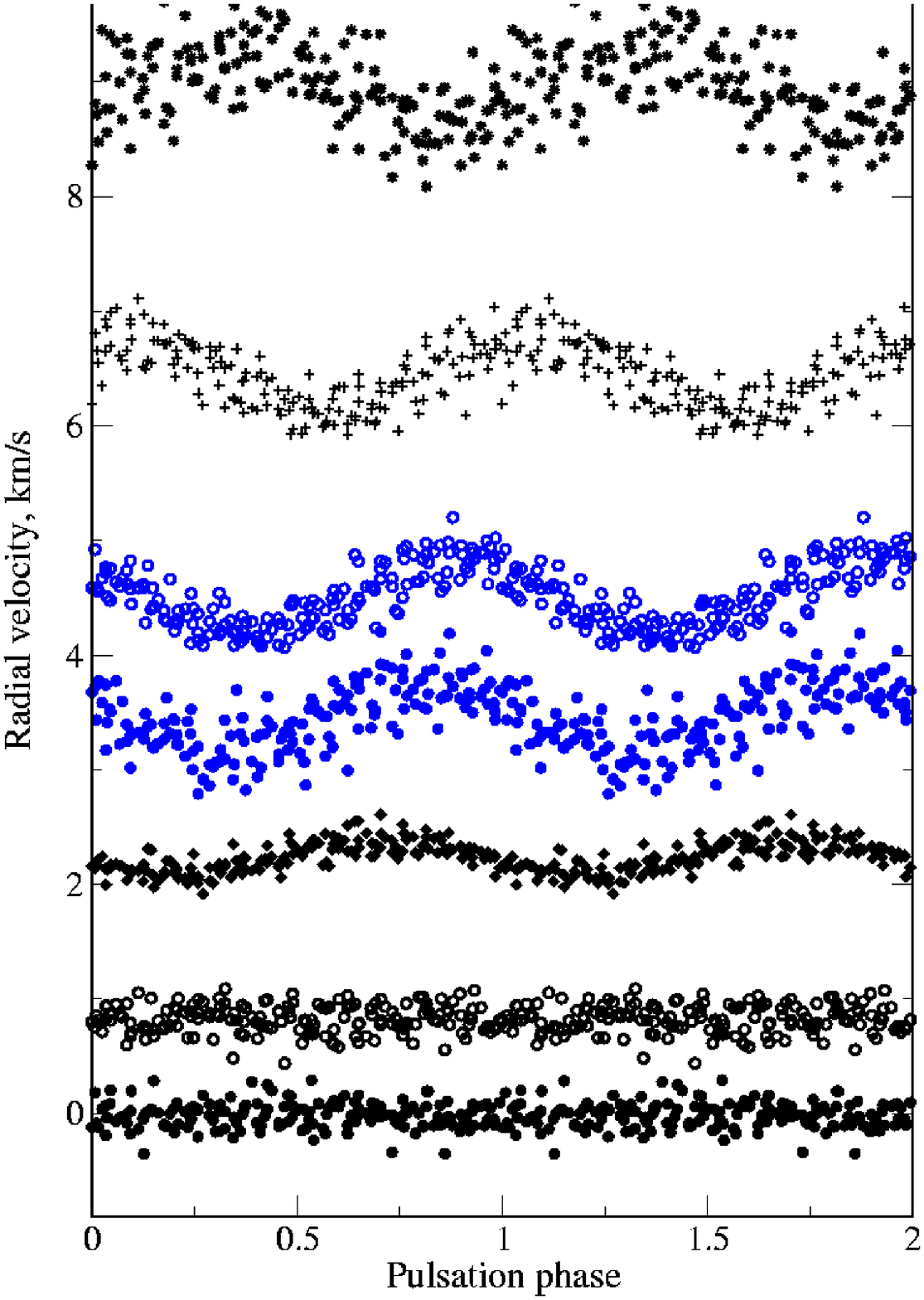}\fifps{8.5cm}{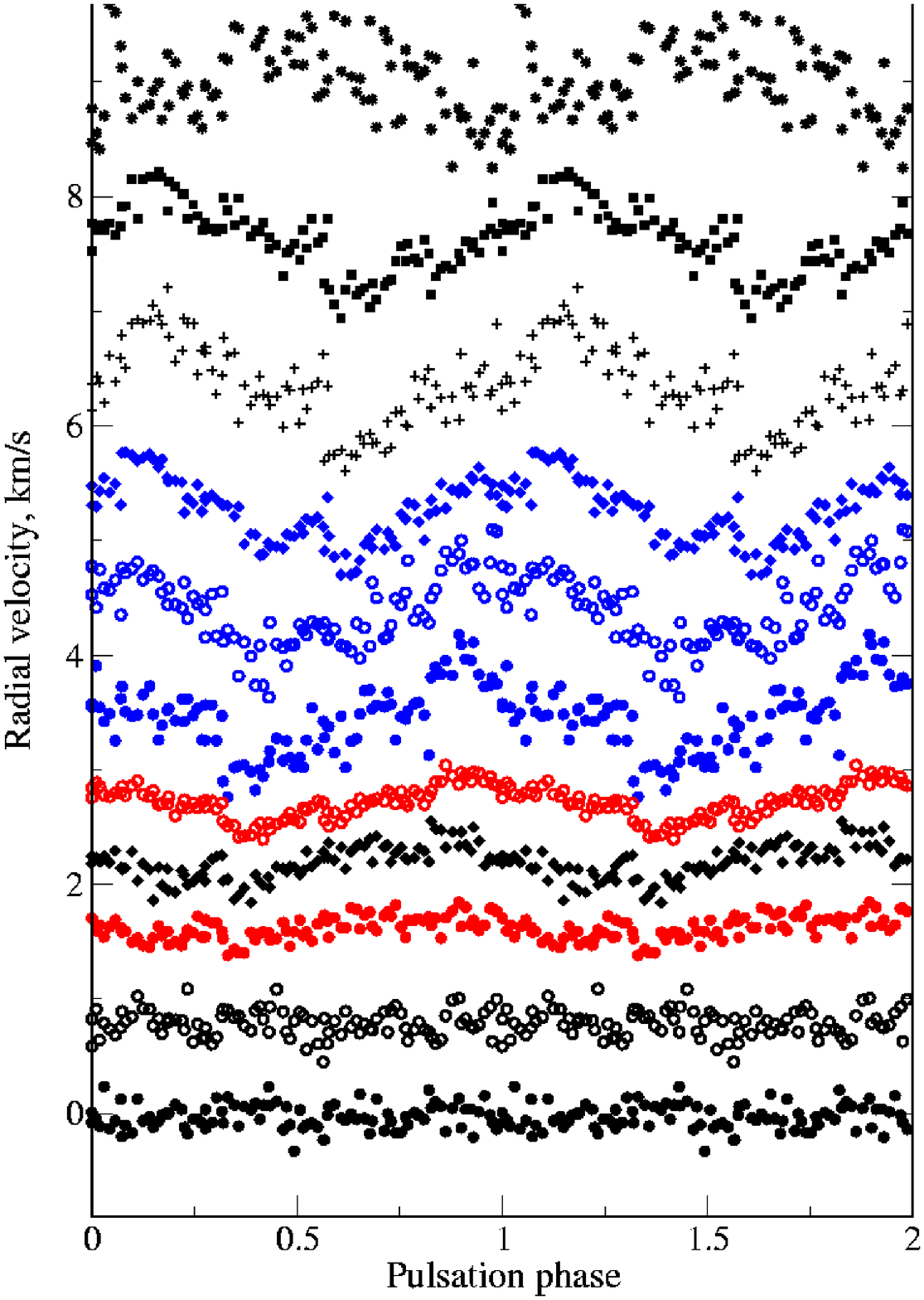}
\caption{Radial velocity variations with a period of P\,=\,6.125\,min for selected spectral lines in 2001 (left panel) and in 
2004 (right panel). In the right panel lines from bottom to top are: Ca\i\,6717, Fe\i\,6678, H$\beta$ (core), Eu\ii\,6645, 
H$\alpha$ (core), Nd\ii\,6650, Nd\iii\,6690, Nd\iii\,6550, Pr\iii\,6706, Er\iii\,4735, and Tb\iii\,6688 \AA. Same lines except of 
H$\beta$ (core), H$\alpha$ (core), Nd\iii\,6550, and Er\iii\,4735 are displayed in the left panel.} 
\label{RV}
\end{figure*}

\section{Discussion}             
\label{sec:discuss}

We have obtained time-resolved observations with different spectrographs in different years (2001, 2003 and 2004) and at
nearly the same rotation phases close to the magnetic maximum. We have already reported the similarity of phase shifts
between RV variations in the lines of different elements/ions (Sachkov~et~al. \cite{sach04b}) based on the observations taken in 
2001 and 2003. Figure\,\ref{RV} gives another comparison of the RV variations derived from the observations taken in 2001 
(GECKO) and in 2004 (HARPS) at rotation phase 0.87. 
It was mentioned in Sect.\,\ref{freq} that the frequency analysis of our observations in 2001 and 2004 reveals the same 
highest amplitude frequencies of 2720.9 and 2652.9\,$\mu$Hz, 
which corresponds to pulsation periods of 6.125 and 6.28\,min. In 2003 we got only 53 spectra and hence we can not resolve frequencies,
therefore a mean period of 6.20\,min was used. RV amplitudes and phases for a sample of common lines observed in 2001, 2003 and 2004 are given
in Table\,\ref{tbl5} (Online material). Phases in 2001 and 2003 were calculated relative to the HJD of the first observation in a given year and were shifted for comparison purpose by $-0.1$ and $-0.2$, respectively. 
Obviously, the same lines are variable and the similarity of amplitudes and 
phases of the RV maximum indicate stability of the pulsation pattern in the atmosphere of \hd\ at least during recent years.

Simultaneous photometry and spectroscopy allow us for the first time to 
phase accurately RV variations due to pulsation observed in different spectral lines with the photometric pulsation signature. 
To determine a phase shift between RV and light variations we used the pulsation frequency with the largest photometric amplitude 
in the WET and MOST data (6.125\,min).

Our results show a gradual decrease of the phase lag from the H$\beta$ line to Tb\iii\ lines, which may be interpreted as a 
running wave in the atmosphere of \hd, if different lines are formed at different atmospheric layers. The same phenomenon, 
known as the Van Hoof effect, was found earlier in $\beta$\,Cep-type stars (Mathias \& Gillet \cite{bcep}) and allowed them 
to derive the propagation time of the running wave through the stellar atmosphere. 
A first estimate of the running pulsation wave speed in the atmosphere of \hd\ can be obtained from the phase lags and the respective 
formation depths of the Nd\ii\ and Nd\iii\ lines determined according to Mashonkina~et~al. (\cite{MRR05}). In the relevant atmospheric layers 
($-6.2\le\log\tau_{5000}\le-4.2$) the pulsation wave propagates with nearly constant speed of $\sim$6~\kms\ which 
is slightly less than the sound speed in adiabatic approximation.

An analysis of the running wave properties in a roAp star atmosphere is rather difficult. First, we know that
elements are stratified in the atmospheres of Ap and roAp stars (Babel \cite{Babel92}; Ryabchikova~et~al. \cite{RPK02},
\cite{RLK05}), and therefore stratification has to be taken into account for the line formation depth calculations.
Second, a stratification analysis of the REEs, which are the main carriers of pulsation information, is not correct
without considering NLTE effects (Mashonkina~et~al. \cite{MRR05}). The third important issue is the surface
inhomogeneity in the chemical composition. 
\hd\ is a spectrum variable and the first rough analysis of the element
distribution was published by Preston (\cite{Preston}). Later, Ryabchikova~et~al. (\cite{RTMS00}) showed evidence for a 
concentration of the Fe-peak elements in a wide band around the magnetic equator,
while REEs (in particular Pr and Nd) and Co are concentrated in large spots near, but not exactly at one magnetic pole 
(the other pole is never visible). Our magnetic field measurements support this difference in the element surface distribution.

Recent Magnetic Doppler Imaging of \hd\ (L\"uftinger~at~el. \cite{MDI06}) revealed a small but non 
negligible difference (both in longitude and latitude) between the surface distributions of different REE elements.    
Therefore, part of the phase shifts between RV curves {\em may} be due to the different chemical surface distribution relative to 
the magnetic pole. Cunha (\cite{cunha06}) showed that phase shifts may reach 0.25 of the period between the magnetic pole and 
the magnetic co-latitude $\theta\sim30\degr$ (see lower panel of Fig.\,6 in Cunha \cite{cunha06}).  
  
\begin{figure}[!th]
\centering
\firps{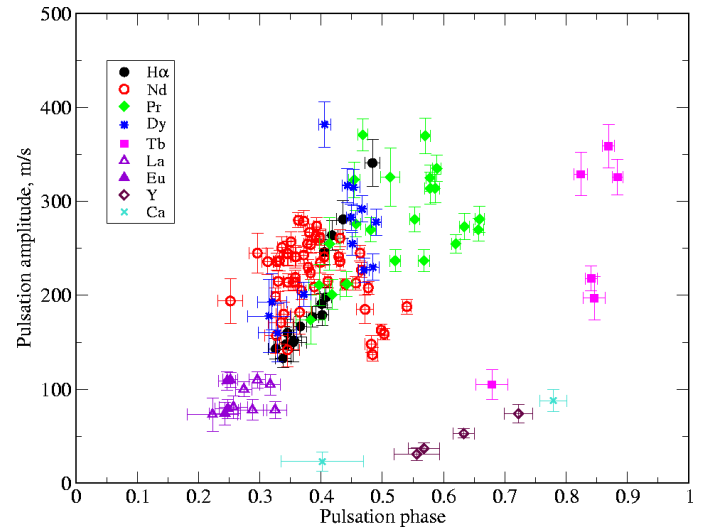}
\caption{Pulsation amplitude of spectral lines versus pulsation phase at RV maximum, based on a pulsation period of 6.125\,min.}
\label{uves1_rv-ph}
\end{figure}

A detailed study of the line formation depth in the atmosphere of \hd\ will be subject of a forthcoming paper. However, 
some first information may be obtained already without such a detailed analysis when plotting the maximum RV value 
observed for a given spectral line as a function of pulsation phase, determined from photometry. If the photometric 
pulsation phase of the maximum RV is related to specific line-forming layers in the stellar atmosphere, then we expect 
new insights in a pulsating atmosphere. Figure\,\ref{uves1_rv-ph} shows RV amplitudes as a function of pulsation phases 
for numerous lines of several elements measured in UVES spectra (rotation phase 0.944). These pulsation data include 
bisector measurements of the H$\alpha$ core and the cores of Ca\i\ and Ca\ii\ resonance lines.

The data seem to cluster along two curves separated roughly by 0.5 of a pulsation period (Fig.\,\ref{uves1_rv-ph}). The
first curve may be directly connected with optical depth, thanks to NLTE calculations for hydrogen and Nd lines
(Mashonkina~et~al. \cite{MRR05}; Sachkov~et~al. \cite{Rome}). The RV amplitude grows towards the upper layers, reaches 
a maximum and then decreases, as indicated by the Nd and Pr lines. 

At present it is difficult to conclude if the second curve represents another part of a continuous amplitude--phase distribution, 
because it is defined mainly by Y and Tb lines, which formation depths are unknown. We think that Tb lines are formed at about 
the same depth or higher than Nd, Pr and other REE lines. The core of the resonance Ca\ii\,3933\,\AA\ line (pulsation phase of 0.8) 
may be also formed high in the atmosphere, particularly if one considers possible stratification of Ca. But the Y lines are probably 
formed in lower atmospheric layers.

Our observations allow us to check the claimed pulsation phase jumps of 180$\degr$ at rotation phases corresponding to magnetic 
extrema (Mkrtichian \& Hatzes \cite{mkrt05}). These authors had to link two sets of observations, separated by one month 
(more than 9000 pulsation cycles), to cover these phases. The phase jumps reported by Mkrtichian \& Hatzes occur exactly between 
the two sets of observations. We, on the other hand, do not observe any phase changes exceeding 35$\degr$ (0.1 of the pulsation period) 
in our data sets, which fortunately happen to cover in 2002 the magnetic minimum in 4 consecutive nights (sets 4 to 10) 
and in 2004 the magnetic maximum in 4 nights of which three were consecutive (sets 12 to 15). Thus,
we believe that the rotational modulation of pulsation phase reported for \hd\ by Mkrtichian \& Hatzes 
(\cite{mkrt05}) is spurious.

Finally we want to point again to Table\,\ref{puls} which provides the basis 
for the present investigation. It is a compilation of 
unblended spectral lines measured in the spectral range from $\lambda$\,3900 
to 6800\,\AA\ and of further 80 lines bluewards the Balmer jump. Because of its volume 
this table is available only as on-line material. We present the measured central 
wavelengths of nearly 600 spectral lines in \AA, followed by the pulsation amplitude 
in \ms\ and the amplitude error, the period with the largest amplitude determined 
with a least-squares fit after a periodogram analysis and the probability of the given 
period (Horne \& Baliunas \cite{horne}). The next four columns give amplitude, amplitude error, phase relative to the main 
photometric period observed by MOST (6.125\,min) and the phase error. These data are followed 
by the same information, but relative to the third prominent photometric period observed by 
MOST (6.282\,min), which is the second prominent spectroscopic period. And finally we give in the last column additional information.
This table includes information of pulsation properties of $\sim$600 
spectral lines in the roAp star \hd\ observed in 2004 in a range of more than 3000\,\AA. 
Because of the known rotational modulation of the pulsation RV amplitudes (see Fig.\,\ref{shift}) the lines measured in both observing runs have different amplitudes in Table \,\ref{puls}.

\section{Conclusions}        
\label{sec:concl}

An extensive spectroscopic and polarimetric study of \hd\ provide new information about the pulsation properties of a
roAp star atmosphere. With this new analysis we confirm our previous results (Sachkov~et~al. \cite{sach04b}) and of Balona \& Zima 
(\cite{bal_z}), that REE lines and the H$\alpha$ core show large pulsation amplitudes (150 to 400\,\ms), while spectral lines of 
the other elements (Mg, Si, Ca, Fe-peak) are nearly constant. 

Our data permit for the first time to determine directly the phase shifts for different chemical elements, respectively ions, between 
pulsation signatures observed in RV data and in photometry. These shifts, derived from contemporaneous photometric and spectroscopic 
observations, together with magnetic field measurements over a stellar rotation period, will be used in our forthcoming structural 
analysis of the pulsating atmosphere of \hd.

\begin{acknowledgements}
We thank the MOST Science Team for providing us with the
photometric data and frequency analysis of \hd\ prior to publication. This work was supported by the Austrian FFG-ALR 
(MOST Ground Station) and Austrian Science Fund (FWF-P17580N2), by grant 11630102 from the Royal Swedish Academy of Sciences, 
and by the Natural Sciences and Engineering Research Council of Canada. TR and MS acknowledge financial support from RFBR grant 
04-02-16788a and from the RAS Presidium (Program ``Origin and Evolution of Stars and Galaxies''). We also thank our referee, Don W. Kurtz, 
for his constructive comments which helped to improve the paper.
\end{acknowledgements}


\Online

\begin{scriptsize}
\longtab{4}{
\begin{longtable}{c|rrccc|rrcc|rrcc|l}
\caption{\label{puls} Summary of our spectroscopic pulsation analysis of individual spectral lines in \hd. The columns give 
central wavelengths in \AA, followed by the pulsation amplitude $A$ (in \ms), its error $\sigma_A$ 
and the period $P$ (in min) with the respective error estimate $\sigma_P$ derived with a least-squares fit. 
Corresponding probability of periodic signal 
(Prob., calculated according to Horne \& Baliunas \cite{horne}). 
The next two groups of columns give pulsation amplitudes and phases $\phi$ with the respective errors, 
determined with a simultaneous fit of two fixed pulsation periods indicated in the column head.}\\
\hline
\hline
Wavelength &\multicolumn{5}{c|}{Free period} & \multicolumn{8}{c|}{Fixed periods }& Identification \\
\hline
                &         & & & & &\multicolumn{4}{c|}{ 6.125 min } &\multicolumn{4}{c|}{ 6.282 min } & \\
         \AA    & $A$\,(\ms) & $\sigma_A$ & $P~(min)$ & $\sigma_P$ & Prob.& $A$\,(\ms) &$\sigma_A$ & $\phi$ & $\sigma_{\phi}$ & $A$\,(\ms) &$\sigma_A$ &$\phi$&$\sigma_{\phi}$&  \\ 
\hline
\endfirsthead
\caption{Continued.}\\
\hline\hline
Wavelength &\multicolumn{5}{c|}{Free period} & \multicolumn{8}{c|}{Fixed periods }& Identification \\
\hline
                &   & & & & &\multicolumn{4}{c|}{ 6.125 min } &\multicolumn{4}{c|}{ 6.282 min } & \\
                & $A$ & $\sigma_A$ & $P$ & $\sigma_P$ & Prob.& $A$ &$\sigma_A$ & $\phi$ & $\sigma_{\phi}$ & $A$ &$\sigma_A$ &$\phi$&$\sigma_{\phi}$&  \\ 
\hline
\endhead
\hline
\endfoot
\hline
\endfoot
\multicolumn{15}{l}{Hydrogen  } \\
\multicolumn{15}{l}{ HARPS}     \\
  4340.468 &  59 & 15 & 6.237 & 0.041&  0.847&   22 & 18 & 0.132 & 0.132 & 44 & 18 & 0.759 & 0.067&  H$\gamma$ \\
  4861.330 &  91 & 11 & 6.115 & 0.018&  1.000&   93 & 13 & 0.230 & 0.022 & 39 & 13 & 0.605 & 0.053&  H$\beta$ \\
  6562.799 & 175 & 10 & 6.140 & 0.009&  1.000&  168 &  8 & 0.376 & 0.008 & 87 &  8 & 0.785 & 0.015&  H$\alpha$\\
       &     &    &       &      &       &      &    &   &   &    &    &       &      &       \\
\multicolumn{15}{l}{ Na}                                                 \\
\multicolumn{15}{l}{ HARPS}                                                  \\
\hline
\end{longtable}
}
\end{scriptsize}

\newpage
\begin{table*}
\begin{scriptsize}
\caption{Comparison of the pulsational amplitudes and phases near the magnetic maximum at different years calculated with the main
pulsation period P\,=\,6.125\,min in 2001 and 2004, and P\,=\,6.20\,min in 2003.}
\label{tbl5}
\begin{center}
\begin{tabular}{l|cccc|cccc|cccc|cccc|l}
\noalign{\smallskip}
\hline
WL  &\multicolumn{4}{c|}{2001}& \multicolumn{4}{c|}{2003}&  \multicolumn{4}{c|}{2004} &  \multicolumn{4}{c|}{2004}& \\                                                            
\AA &\multicolumn{4}{c|}{phase=0.872}& \multicolumn{4}{c|}{phase=0.083}&  \multicolumn{4}{c|}{phase=0.867} & 
\multicolumn{4}{c|}{phase=0.944}&\\  
\hline
 & $A$ &$\sigma_A$ & $\phi$ & $\sigma_{\phi}$ & $A$ &$\sigma_A$ &$\phi$&$\sigma_{\phi}$& $A$ &$\sigma_A$ & $\phi$ & $\sigma_{\phi}$ &
$A$ &$\sigma_A$ &$\phi$&$\sigma_{\phi}$& \\
\hline
\hline
  5284.690 &	&   &	  &	&319 &37 &0.678  &0.017  & 338 & 11 & 0.685 & 0.006 &259& 11 &0.642 &0.007 &  Pr\iii\			\\
  5299.986 &	&   &	  &	&291 &31 &0.702  &0.016  & 327 & 12 & 0.678 & 0.006 &246& 10 &0.641 &0.007 &  Pr\iii\			\\
  5844.406 &	&   &	  &	&301 &38 &0.571  &0.020  & 305 & 14 & 0.603 & 0.008 &299& 15 &0.576 &0.008 &  Pr\iii\			\\
  5998.935 &	&   &	  &	&254 &30 &0.571  &0.019  & 382 & 18 & 0.606 & 0.008 &329& 15 &0.587 &0.008 &  Pr\iii\			\\
  6160.238 &	&   &	  &	&195 &19 &0.610  &0.016  & 342 & 12 & 0.657 & 0.006 &251& 10 &0.624 &0.006 &  Pr\iii\			\\
\hline	   																			
\end{tabular}
\end{center}
\end{scriptsize}
\end{table*}

\end{document}